# Generative Artificial Intelligence in Qualitative Research Methods: Between Hype and Risks?

Maria Couto Teixeira[0009-0001-1994-0073], Marisa Tschopp[0000-0001-5221-5327], and Anna Jobin[0000-0002-4649-7812]

¹ Human-IST Institute, University of Fribourg, 1700 Fribourg, Switzerland
² Scip AG, Badenerstrasse 623, 8048 Zurich, Switzerland
`anna.jobin@unifr.ch`

**Abstract.** As Artificial Intelligence (AI) is increasingly promoted and used in qualitative research, it also raises profound methodological issues. This position paper critically interrogates the role of generative AI (genAI) in the context of qualitative coding methodologies. Despite widespread hype and claims of efficiency, we propose that genAI is not methodologically valid within qualitative inquiries, and its use risks undermining the robustness and trustworthiness of qualitative research. The lack of meaningful documentation, commercial opacity, and the inherent tendencies of genAI systems to produce incorrect outputs all contribute to weakening methodological rigor. Overall, the balance between risk and benefits does not support the use of genAI in qualitative research, and our position paper cautions researchers to put sound methodology before technological novelty.

## 1    Introducction

This position paper has its beginnings in a research project that explores digital methods [1] and the role of generative AI (genAI) for data analysis in qualitative research. Our reflections are situated within a research project that attempts to critically explore the performativity of Artificial Intelligence over the coming years [2]. Engaging with the growing body of literature on genAI in qualitative research, we wondered whether we would find useful insights on how to use genAI that go beyond the current hype hailing AI as transformative for all and any sector including research [3–5]. However, instead of finding widespread robust new research practices, we identified two salient issues that seemed important but underrepresented.

The first issue concerns the lack of methodological robustness due both to the absence of shared standards and guidelines as well as the black boxed underpinnings of genAI. Indeed, while the literature on genAI in qualitative research in general, and coding in particular, has produced valuable studies and discussions within and across different disciplines, there seems to be no shared understanding of how to apply genAI to qualitative research in practice. Most studies do not report essential details such as model type, parameters, or prompts [6]. Moreover, with commercial platforms remaining closed-source and genAI models evolving at high speed, even if such information were provided, scientific validity would remain severely limited due to the lack of methodological reliability, transparency and rigor.

Compounding this first issue, we identified as a second issue a recurrent tension between potential and risk mentioned in recent studies. Across different study designs and methodological reflections, the insights converge towards lauding genAI for its efficiency and rapidity while simultaneously cautioning against well-known problems such as 'hallucinations', bias, and false negatives.

In light of these persistent shortcomings, our exploration of genAI in qualitative coding has lead us to question it as a possible analytical tool. This paper aims at contributing to the debate about the validity of using genAI, but it does not present a definitive overview of genAI use in qualitative research, nor does it propose guidelines. Instead, it surfaces important but easily overlooked conundrums that may not easily be resolved, if they are solvable at all. Our conclusion calls for a strengthened focus on methodological validity and reliability, which both are fundamental to ethical and trustworthy research.

## 2      Use of GenAI in Qualitative Research

Artificial Intelligence as an umbrella term has gained traction across many domains such as policy, media, and research [7]. Since the release of ChatGPT in 2022, so-called generative AI (genAI) in particular has attracted increased scholarly attention, producing a growing body of studies, not only of genAI itself but also of its use in research [8]. Amidst suggestions to use genAI for topic exploration [9], literature reviews [10], quantitative analyses [11] or simply writing assistance [12], we are particularly interested in qualitative research that includes the method of coding. An inherently subjective, reflective process [13], qualitative coding does not seem predestined to be automated through genAI. However, qualitative research is known to be time-consuming. In a context of tightening resources and high pressure to publish the supposed efficiency of genAI make its use attractive for researchers [14]. Still, use of genAI is not a standardized process, neither in quantitative nor in qualitative research, with many scholars experimenting and testing the boundaries of its limitations [8].

### 2.1      Methodological Robustness and Documentation

In research, the robustness of different methodologies depends on their epistemological claim. For positivist research, for example, reproducibility is key. This means that by using the same methodology and data, researchers should be able to reproduce the results [15]. Interpretivist research, which is often highly context-dependent and can therefore not be reproduced stricto sensu, relies notably on confirmability and dependability [16] to produce valid results. In many academic fields, (post)positivism is the unacknowledged default paradigm. However, positivist criteria cannot coherently be used to assess the quality of interpretivist research. To suggest that genAI can be a technological solution to reduce subjectivity in qualitative analyses [6] therefore indicates a lack of awareness that subjectivity can be viewed and managed differently across paradigms [17]. Even as most qualitative studies are interpretivist, they do not lack rigor but rely on shared standards, for instance regarding research quality [18] or shared procedural guidelines such as grounded theory [19] or thematic analysis [20].

In all empirical research, methods have to be fully documented and explained, including their use of software such as genAI. However, in 2018 already, only 24% of empirical AI studies included full documentation of their methodology [21]. Relating specifically to qualitative coding, a recent mapping found important variances in methodological documentation [6]. Research on the use of genAI in qualitative analyses often mentions the advice of a hybrid use, so-called "human-AI collaboration" [22-25]. What exactly this might look like, however, is unclear, as details of the process remain largely undocumented. Notable exceptions to this general assessment document e.g. the exact model and 'temperature' setting of the genAI used all while acknowledging transparently the subjective nature of the coding process [26]. Others propose a systematic prompt design that follows the development of codebooks in qualitative studies, with humans in the loop to construct prompts in an iterative process [27], or even an entire set of best practices for text annotation with Large Language Models [28]. Still, there is no indication that generalized best practices have been adopted for either the documentation or the use of genAI qualitative research.

## 2.2    Commercial and Technological Opacity

GenAI as a broad term describes diverse techniques and AI models, which are constructed with various parameters such as top-k, the number of tokens or the 'temperature' of the data processing. Outputs are also influenced by the so-called prompts, that is, the queries crafted to elicit such outputs from genAI systems [29].

Much research that includes genAI uses commercial AI systems belonging to for-profit organisations. These organisations do not communicate about the inner workings of their genAI technology, stating security issues and trade secrets [30]. It is therefore not possible for researchers to reliably determine, or even verify, which data had been used to create and train the model, and how the parameters have been set. Such opacity around a central methodological tool is problematic in and by itself.

The commercial opacity is compounded by the technological 'black box' that is AI for individual researchers [31]. Although its outputs appear legible to humans, it is important to note that genAI reportedly lacks understandability [32]. It cannot process ambiguities if there is not enough context-establishing data [33], since it does not actually understand the data it handles. Instead, genAI produces output based on its model's configuration, previous datasets, and probabilities. Text-producing tools such as Chat-GPT, Claude or Perplexity, calculate the probabilities of each word to guess the next ones. While often useful, this is problematic when it results in outputs that appear plausible and correct even if they are not [34].

Such erroneous outputs are often called 'hallucinations' [6], which seems like a misnomer, as all outputs of genAI are in essence confabulated: the truth value of outputs is not inherent in the technological system, but it is assigned through human interpretation [35].

Many studies about genAI use for qualitative coding acknowledge the important limitation of potentially inaccurate output [22, 24, 25]. To mitigate this risk, suggestions range from including humans 'in the loop' along the lines of "human-AI collaboration" [24] to always verifying genAI output [36]. However, like with use of genAI in general, there are no shared best practices of how to verify genAI outputs in particular. Moreover, it is unclear to understand what verification would look like considering the lack of documentation and technical information available.

Overall, the inherent technicalities of how genAI works, including its tendency to introduce 'hallucinations' and reproduce bias, are contrary to qualitative validity that relies on robust, reliable methods and transparency. Such limitations therefore severely impact methodological robustness and validity more broadly, and open the larger issue of how to balance them against the often-repeated potential of genAI for qualitative research.

## 3      Balancing Potentials and Risks of using GenAI

One of the recurring potential benefits of using genAI for qualitative coding methods is said to be efficiency [14]. Coding as an activity of pattern recognition is stated as a task that genAI could do much faster than human coders [25]. Some scholarship explicitly links the supposed efficiency of genAI to gains in time and, thus, money [24].

Statements about the benefit of genAI efficiency in coding, however, fail to take into account the time required for documenting genAI use and verifying genAI outputs (not even counting the initial effort of learning how to use it and staying up to date with the technology's constant changes) [37, 29]. As explained above, both of these activities are indispensable for a sound methodological approach to qualitative research with genAI. On top of that, verification becomes even more time-consuming when such verification not only consists of checking whether genAI output is correct (identifying true positives vs. false positives), but also searching for potentially missing output (identifying true negative vs. false negative) [38, 39]. It is doubtful that genAI can still be considered to be efficient when every output has to be thoroughly verified.

In addition, there are other risks to be considered when using genAI for qualitative coding. For instance, biases in the training data can lead to biased codebooks or skewed analyses [36].  There is also the risk of a potential future lack of codebook diversity [40]. For one, genAI suggesting codes to researchers could limit the creativity of their coding process, as beginners and even experts may start to overly depend on such suggestions [41]. Additionally, over time, genAI might lead to "data contamination" [28] by computing new codes or codebooks based on existing codes or codebooks, resulting in an ever limited scope of possible output. Instead of creative iterations by human coders, automated streamlining can resulting in poorer coding diversity and quality [42].

Many of these reflect problematic issues already identified more generally regarding AI [43]. In certain sectors, standardization attempts to deal with AI risks are underway, notably triggered by the EU AI Act [44]. However, at present, no standardized guidelines for qualitative coding exist that make genAI methods meet the research criteria for dependability and confirmability [16].

The risks adversely impacting the quality and ethics of a qualitative inquiry, the lack of canonical guidelines, and the challenges in providing sufficient documentation when making use of genAI in qualitative coding, partly reinforce one another. 'Hallucinations', missing information and biases in qualitative coding with genAI cannot be dealt with systematically as there are no shared methods to assess and mitigate these risks. Because no methodological guidelines exist, there is no standardized way of documenting genAI use. Then again, tthe lack of documentation makes it hard to provide a foundation for critically assessing and mitigating risks. Taken together, these aspects converge toward a call for great hesitation about integrating genAI into qualitative research methods.

Last but not least, this is also an ethical issue. Research ethics generally requires risks to be outweighed by benefits [45]. Considering the important risks, the use of genAI should therefore come with great benefits. However, as discussed above, the most prominent benefit mentioned is time savings. When considering the fact that all genAI output must be checked for both errors and incompleteness, the importance of this supposed benefit diminishes significantly. Given the technical and ethical limitations of genAI, it is legitimate to interrogate whether genAI should be used in qualitative research at all.

## 4     Conclusion

This position paper explores the use of genAI in qualitative coding with the aim of providing a baseline for a much-needed scholarly debate about its validity. We argue that, in its current opaque form, genAI cannot meet qualitative research criteria such as confirmability, dependability, and transparency. Outputs of genAI systems are not inherently correct or trustworthy, and certainly not more trustworthy than those achieved through the labor of experienced researchers. The supposed efficiency and productivity gain of genAI is counteracted by the burden to verify all outputs for 'hallucinations', that is, false positives, and to account for false negatives, that is, wrongfully omitted output. Without robust methodological standards, using genAI introduces more risk than benefits.

We therefore argue that academic researchers should not focus on the use of genAI for AI's sake. Robust methodology must be prioritized over technological novelty, because empirical research is based on trustworthy, valid inquiries. Following the current AI hype by uncritically adding genAI to otherwise proven qualitative research methods risks producing and even normalizing invalid research practices.

**Acknowledgments.** This research was funded in part by the Swiss National Science Foundation (SNSF), grant no. 10002211. The authors are grateful to the reviewers for their helpful feedback, especially reviewer 2's epistemological precisions.